\documentclass[useAMS]{mn2e}
\usepackage{epsfig}
\DeclareMathSymbol{\gtrsim}       {\mathrel}{AMSa}{"26}
\DeclareMathSymbol{\lesssim}      {\mathrel}{AMSa}{"2E}
\newcommand{\be}{\begin{equation}}
\newcommand{\ee}{\end{equation}}
\newcommand{\kr}{\kappa_{\rm R}}

\newcommand{\rmd}{\,{\rm d}}
\newcommand{\rcz}{R_{\rm CZ}}
\newcommand{\Ne}{N_{\rm e}}
\newcommand{\etal}{et al. }
\newcommand{\rms}{{\rm s}}
\newcommand{\rme}{{\rm e}}
\newcommand{\rmion}{X({\tt NE})}
\title[Up-dated opacities from the Opacity Project]
{Up-dated opacities from the Opacity Project}
\author[N. R. Badnell \etal ]
{N. R. Badnell$^1$, 
M. A. Bautista$^2$, 
K. Butler$^3$,
F. Delahaye$^{4,5}$, 
C. Mendoza$^2$,
\newauthor 
P. Palmeri$^6$\thanks{Present address: Astrophysique et Spectroscopie, 
Universit\'e de Mons-Hainaut, 20 Place du Parc, B-7000 Mons, Belgium},  
C. J. Zeippen$^5$
and M. J. Seaton$^7$\\
$^1$Department of Physics, University of Strathclyde, Glasgow, G4 0NG\\
$^2$Centro de F\'{\i}sica, IVIC, PO Box 21827, Caracas 1020A, Venezuela\\
$^3$Universit\"{a}tssternwarte M\"{u}nchen, Scheinerstra\ss e 1, D-81679, M\"{u}nchen, Germany\\
$^4$Department of Astronomy, Ohio State University, Ohio, 43210-1173, USA\\
$^5$LUTH, Observatoire de Paris, F-92195, Meudon, France\\
$^6$NASA Goddard Space Flight Center, Code 662, Greenbelt, Maryland, 20771, USA\\
$^7$Department of Physics and Astronomy, University College London,
London, WC1E 6BT}
\begin{document}
\date{Accepted XXX. Received XXX; in original form XXX}

\pagerange{\pageref{firstpage}--\pageref{lastpage}} \pubyear{2004}

\maketitle

\label{firstpage}

\begin{abstract}
Using the code {\sc autostructure}, extensive
calculations of inner-shell atomic data have been made 
for the chemical elements He, C, N, O,
Ne, Na, Mg, Al, Si, S, Ar, Ca, Cr, Mn, Fe and Ni. The results are
used to obtain up-dated opacities from the Opacity Project, OP. 
A number of other improvements on earlier work have also been included.

Rosseland-mean opacities from OP are compared 
with those from OPAL. Differences of 5 to 10\% occur.
OP gives the `$Z$-bump', at $\log(T)\simeq 5.2$, to be shifted to 
slightly higher temperatures. The opacities from OP, as functions of
temperature and density, are smoother than those from OPAL.

The accuracy of the integrations used to obtain mean opacities 
can depend on the frequency-mesh used. Tests
involving variation of the numbers of frequency points show that for
typical chemical mixtures the OP integrations are numerically correct to
within 0.1\%. 

The accuracy of the interpolations used to obtain mean opacities
for any required values of temperature and density depend on the
temperature--density meshes used.
Extensive tests show that, for all
cases of practical interest, the OP interpolations give results
correct to better than 1\%.

Prior to a number of recent investigations which have indicated a need 
for downward
revisions in the solar abundances of oxygen and other elements,
there was good agreement between properties of the sun
deduced from helioseismology and from stellar evolution models calculated
using OPAL opacities. The revisions destroy that agreement. In a 
recent paper Bahcall \etal argue that the agreement would be restored
if opacities for the regions of the sun with 
\mbox{$2\times 10^6 \lesssim T \lesssim 5\times 10^6$ K} ($0.7$ to $0.4 R_{\sun}$) 
were larger than those given by OPAL by about 10\%. In the region concerned,
the present results from OP do not differ from those of OPAL by more
than 2.5\%.

\end{abstract}
\section{Introduction}
First results for opacities from the Opacity Project (OP) were
published by Seaton \etal (1994, to be referred to as Paper I)
and were in good general 
agreement with those from the OPAL project (Rogers and Iglesias, 1992)
except for regions of high densities and temperatures. In a more
recent paper, Badnell and Seaton (2003, Paper II) confirmed a suggestion
made by Iglesias and Rogers (1995) that the discrepancies in
those regions were due to the omission of important inner-shell
processes in the OP work. Further comparisons between OPAL results
and those from OP with inclusion of all important inner-shell
processes are given by Seaton and Badnell (2004, Paper III) for the case
of the 6-element mix (H, He, C, O, S and Fe) of Iglesias and Rogers (1995). 
In Papers II and III, the required inner-shell atomic data were computed using
the code {\sc autostructure} (Badnell, 1986, 1997). Similar 
computations of the required inner-shell
atomic data for all cosmically-abundant elements % \marginpar{$\star$}
(He, C, N, O, Ne, Na, Mg, Al, Si, S, Ar, Ca, Cr, Mn, Fe and Ni) have
now been made. Some further improvements and up-dates of the OP work are
also considered in the present paper.

\section{Outer-shell atomic data} 
Most of the original OP outer-shell atomic data were computed using
sophisticated R-matrix methods (see The Opacity Project Team, 1995, 1997).
Ions are specified by nuclear charge $Z$ and number of
`target' electrons {\tt NE}: the total number of electrons in an ion
is equal to ({\tt NE}+1). For iron ions with {\tt NE}$=13$ to 18, the R-matrix
data were supplemented by extensive data calculated 
by Lynas-Gray \etal (1995)
using the code {\sc superstructure} (Eissner \etal, 1974). Data for
the less-abundant iron group elements, Cr, Mn and Ni, were obtained from iron
data using interpolation techniques described in Paper I.

All original outer-shell atomic data were computed in LS coupling
and allowance for fine-structure was included using methods
described in Paper I.

In Paper III the experiment was made of replacing all of the original iron 
data for {\tt NE } $=13$ to 18
by data computed in intermediate coupling  using
{\sc autostructure}. Use of those new data did not make any
major changes, which provided a good check on the earlier
work, but there were some improvements due to the inclusion of
intercombination lines and some improvements in photoionization
cross-sections. The new intermediate-coupling iron data for iron ({\tt NE } $=13$
to 18) are used in the present work.

As in Paper I, we use data from Kurucz (1988) for the first few
ionization stages of iron-group elements.
\section{Inner-shell atomic data}
All inner-shell atomic data were calculated using {\sc autostructure}
employing  both Russell--Saunders coupling (LS) and intermediate coupling (IC). 
In Paper II it was reported that, for the inner-shell work, the use of LS 
data and  allowance for fine-structure
using methods described in Paper I, did not give opacities
differing significantly from those obtained using the IC atomic
data. The LS inner-shell data were used in the present opacity
work.

New inner-shell data have been calculated for the elements 
N, Ne, Na, Mg, Al, Si, Ar, Ca, Cr, Mn and Ni for {\tt NE} $=1 - 12$.
The present data were generated by working iso-electronically.
Data for He, C, O, S and Fe were reported-on in Paper II
and were generated by working iso-nuclearly. For C, O and S, this
has led to the calculation of some additional inner-shell data (see below).

The K-shell processes that were included are of the form:
\be 1\rms^q 2l^p n'l' +h\nu\rightarrow 1\rms^{q-1} 2l^p n'l' +\rme^- 
\label{K2}
\ee
for photoionization and
\be 1\rms^q2l^pn'l'+h\nu\rightarrow 1\rms^{q-1}2 l^pn'l'n''l''
\rightarrow \rmion + \rme^-
\label{K1}
\ee
for photoexcitation (--autoionization), where
$2l^p$ stands for 2s$^s$2p$^t$ with $p=s+t$. 
Calculations were made for $q=1$ for $p=0$ and $q=2$ for $p=0-7$. 
We used $n', n'' = 2$ to 5, for all allowed $l',l''$.
The contributions from higher-$n$ in (\ref{K1}) were obtained by matching onto
the results of (\ref{K2}).

The L-shell processes that were included are
\be 
2l^q 3l'^p n''l'' +h\nu  \rightarrow 2l^{q-1} 3l'^p n''l''+\rme^-
\label{L2} 
\ee
and
\be 
2l^q 3l'^p n''l'' +h\nu \rightarrow 2l^{q-1} 3l'^p n''l'' n'''l'''
\rightarrow \rmion + \rme^-,
\label{L1} 
\ee
where  $3l'^p$ stands for for
3s$^s$3p$^t$3d$^u$ with $p=s+t+u$. Calculations were made for
$q=1-7$  for $p=0$ and $q=8$ for $p=0-2$.
We used $n', n'', n''' = 3$ to 6 for photoexcitation and $= 3$ to 5 for 
photoionization, for all allowed $l',l'',l'''$.

The M-shell processes that were included are
\be 
3l^q n'l' + h\nu \rightarrow 3l^{q-1} n'l' +\rme^-
\label{M2} 
\ee
and
\be
3l^qn'l'+h\nu\rightarrow3l^{q-1}n'l'n''l''
\rightarrow \rmion + \rme^-.
\label{M1}
\ee
Calculations were made for $q=1$ and 2 and $n', n'' = 3$ to 6.
Calculations for $q=3$ ({\tt NE=13}) for Fe were made previously, and
reported-on in Paper II.

The range of $p$ and $q$ used in equations (\ref{K2}) -- (\ref{M1}) are the
same as were used in Paper II for Fe, while those used previously for C, O and S
were more restrictive (but sufficient for the 6-element mix).
The same range of $p$ and $q$ has now been used for all elements.
Further details of the calculations with {\sc autostructure} may be found in Paper II.

The calculation using {\sc autostructure} provided totals of $11\,523\,624$
spectrum lines and $815\,989$ photo-ionization cross-sections for use
in the opacity work.

\section{The frequency mesh}
\label{mesh}
Let $u=h\nu/(kT)$ and let $\sigma(u)$ be the monochromatic
opacity cross-section per atom. The Rosseland-mean cross-section
is $\sigma_{\rm R}$ where
\be \frac{1}{\sigma_{\rm R}}=\int_0^\infty \frac{1}{\sigma(u)} 
F(u) \rmd u \label{F1}\ee
and
\be F(u)=[15/(4\pi^4)]u^4\exp(-u)/[1-\exp(-u)]^2.\label{F2}\ee
The Rosseland-mean opacity per unit mass is $\kr=\sigma_{\rm R}/\mu$
where $\mu$ is the mean atomic weight.

The function $F(u)$ has a maximum value for $u=3.8300\ldots$ and is small
for $u$ small and $u$ large. It is not practicable to use
a mesh in $u$ such as to resolve all spectral-line profiles, 
the number of points would  be much too large. We therefore
seek a mesh such that trapezoidal-rule integrations give
results of acceptable accuracy. In Paper I a mesh was used
with $10^4$ points in the range $0.001 \le u \le 20 $.

Equation (\ref{F1}) may be replaced by
\be \frac{1}{\sigma_{\rm R}}=\int _{v=0}^{v_{\rm max}} 
\frac{1}{\sigma(u)} \rmd v\label{F3}\ee
where
\be v(u)=\int_0^v F(u) \rmd u \label{F4}\ee
and $v_{\rm max}=v(u \rightarrow \infty)$. From numerical
integration, $v_{\rm max}=1.0553\ldots$. With a given total
number of integration points, {\tt NTOT}, and constant intervals in
$u$ or $v$, use of (\ref{F3}) in place of (\ref{F1})  gives more points in 
regions giving large contributions to the integrals. It was
found that the use of (\ref{F3}) gave a significant improvement in the
accuracy of calculated Rosseland-means.
\section{Element abundances}

Results of Paper I were for a solar mix referred to as S92,
based on work of Anders and Grevesse (1989) together with
some later revisions. Abundances used by Bahcall and Pinsonneault
(2004) in constructing a solar model, which we will refer to as
BP04, are not very different from those of
S92 or of GN93 (Grevesse and Noels 1993) used in much of the OPAL work.
However, recent work (Asplund \etal
2004, and references cited therein) has suggested a need for
substantial revisions in solar abundances for oxygen and other 
elements. Dr N. Grevesse has kindly provided us with an
updated table of recommended abundances, which we will refer
to as A04, taking account of recent work
by Asplund and others. Table 1 gives,
for the 17 elements considered in the present work, the abundances
from BP04 and A04: BP04 gives  $X=0.7394$, $Z=0.0170$ while
A04 gives $X=0.7403$, $Z=0.0123$.

A detailed discussion of recent work on the solar chemical composition
is given by Asplund, Grevesse and Sauval (2005).
\begin{table}
\caption{Element abundances, $A(n)=\log(N(n))$, relative to
$A($H$)=12$.}
\begin{center}
\begin{tabular}{ccccccccc}
\hline
Elem. &  BP04  & A04   & \,\,\, & Elem. & BP04 & A04 \\%[3mm]
\hline
H     &  12.00 & 12.00\,\,\, &        & Si    & 7.56 & 7.51\\
He    &  10.92 & 10.925 &       & S     & 7.20 & 7.14\\
C     &   8.52 & 8.41  &        & Ar    & 6.40 & 6.18\\ 
N     &   7.92 & 7.80  &        & Ca    & 6.35 & 6.31\\
O     &   8.83 & 8.66  &        & Cr    & 5.69 & 5.62\\
Ne    &   8.08 & 7.84  &        & Mn    & 5.53 & 5.46 \\
Na    &   6.32 & 6.17  &        & Fe    & 7.50 & 7.44\\
Mg    &   7.58 & 7.55  &        & Ni    & 6.25 & 6.18\\
Al    &   6.49 & 6.37  &        &       &      & \\
\hline
\end{tabular}
\end{center}
\end{table}
\section{Contributions of inner-shell transitions to Rosseland-means}
We use the variable
\be R=\rho/T_6^3 \ee
introduced in the OPAL work, where $\rho$ is mass--density and
$T_6=10^{-6}\times T$ with $T$ in K. The behaviour of $R$
in some stellar models is shown on Figs. 1 and 2 of Paper
I.

Fig. 1 shows, for the S92 solar mix,  $\log(\kr)$ against
$\log(T)$ for $\log(R)=-1, -2, -3, -4, -5$ and $-6$, both with and
without inclusion of inner-shell transitions. It is seen that
the inner-shell transitions contribute very little to the means
at the lower densities, say $\log(R)\lesssim -4$. The reason is that,
at a lower density, ions in a given ionization stage will 
have maximum abundance only at a lower temperature: and at 
the lower temperatures features in the monochromatic opacities
due to inner-shell transitions 
will appear at larger values of the frequency-variable $u$ 
corresponding to regions where the weighting function $F(u)$ in
equation (\ref{F1}) is small.
\section{Comparisons with OPAL}
OPAL data can be obtained from the OPAL
web site\footnote{www-phys.llnl.gov/Research/OPAL/} for any
required composition. The basic data are given as tables
of $\log(\kr)$ as functions of $\log(T)$ and $\log(R)$. Codes
are given for interpolations to any required values of $\log(T)$ and
$\log(\rho)$ and further interpolations in $X$ and $Z$. 

The OP calculations of monochromatic opacities for each
chemical element are made on a mesh of values of $\log(T)$ and
$\log(\Ne)$ where $\Ne$ is electron density (the meshes
used will be discussed further in Section \ref{chek}). A code {\sc mixv.f}
reads the monochromatic opacities and calculates Rosseland-means
on the OP $(T,\Ne)$ mesh. A further code, {\sc opfit.f}, provides
interpolations to any required values of $T$ and $\rho$. It includes
a facility to produce tables in OPAL format. Such tables will be
used in comparing results from OPAL and OP.

Fig. 2 shows $\log(\kr)$ against $\log(T)$ for %\marginpar{$\star$}
$\log(R)=-1, -2, -3, -4 $ and $-6$, from both OP and OPAL. The
over-all agreement is seen to be good. The feature at $\log(T)\simeq 5.2$,
usually known as the `$Z$-bump', is mainly due to transitions
in iron ions with {\tt NE}$=13$ to 18. It is seen that, compared to
OPAL, the OP feature is shifted to slightly higher temperatures.

Use of the logarithmic scale for $\log(\kr)$, as in Fig. 2,
does not allow one to see the finer details of the level of
agreement between the two calculations. Percentage differences,
(OP--OPAL), are shown on Fig. 3 for $\log(R)=-1.0$ to $-2.5$, and
on Fig. 4 for $\log(R)=-3.0$ to $-4.5$. It is seen that the general
level of agreement is in the region of 5 to 10\%. An excursion 
to larger differences at $\log(T)\simeq 5.5$, approaching 30\% at   
$\log(R)=-4.5$, is due to the differences in the high-temperature
wing of the $Z$-bump, as shown on Fig. 2.

The differences of Figs. 3 and 4 have a somewhat ragged appearance
which results from some lack of smoothness in the values of
$\kr$ at the level of 1 or 2\%. Fig. 5 shows
$\partial\log(\kr)/\partial\log(T)$ at constant $\log(R)$, calculated
using first differences, for $\log(R)=-2$. The data from OP are
seen to be smoother than those from OPAL. Use of the $v$-mesh in
the OP work (see Section \ref{mesh}) gives an improvement in the smoothness.
\section{The solar radiative interior}
Helioseismology provides remarkably accurate values for the depth,
$\rcz$, of the solar convection zone (CZ). 
With the  earlier estimates of solar element abundances, such as S92 
 or GN93 (see section 6), values 
of $\rcz$ and other data from helioseismology were found to be in good
agreement with results from solar models computed using OPAL opacities.
However, with the
new abundances it is found in  two recent papers to be necessary to
increase opacities in the vicinity of $\rcz$: by 19\% according to
Basu and Antia (2004); and by 21\% according to Bahcall \etal
(2004a).  In a further  detailed study of the problem, Bahcall \etal
(2004b) show that there are also discrepancies for the solar profiles
of sound speed and density and of helium abundance. They argue
that all such discrepancies would be removed if the opacities were 
larger than those from OPAL by about 10\% in the region of 
$2\times 10^6$K $\lesssim T \lesssim 5\times 10^6$K 
($0.7$ to $0.4 R_{\sun}$).

In the solar model of Bahcall and Pinsonneault (2004), referred to
as BP04, the base of the convection zone is at $\log(T)\simeq 6.34$ 
($r\simeq 0.715 R_{\sun}$). In the CZ the abundances are independent 
of depth
and the values used for the model were similar to those of 
the S92 mix (i.e. not taking account of recently proposed revisions).
Below the convection zone, $\log(T)\gtrsim 6.34$ 
($r\lesssim 0.715 R_{\sun}$), abundances depend
on the nuclear reactions which have taken place and vary with depth.
Fig. 6 shows the Rosseland-mean opacities adopted by BP04,
using OPAL data. In the CZ the exact values of
opacity are not of great importance for the calculation of
a model, since in that region the temperature gradient is
no longer controlled by the opacity, but comparisons of 
different opacity calculations is still of interest.

Fig. 7 shows percentage differences, (OP--OPAL), between opacities
from OP and those used for BP04. In the region below $\rcz$ the
relative abundances of metals are assumed to be independent of 
depth but the mass-fractions $X$, $Y$ and $Z$ are variable. 
We use a code {\sc mixz.f} to obtain monochromatic opacities for
the specified mixture of metals, and a code {\sc mxz.f} for
variable $X$ and $Z$ (with $Y=1-X-Z$).
At the lowest
temperatures (say $\log(T)\leq 4.5$) $\kr$ is varying very
rapidly as a function of $T$ (see Fig. 6) and the differences
(OP--OPAL) are as large as 10\%. Throughout the rest of the CZ
the differences are never larger than about 5\%. 

\begin{table}
\begin{center}
\caption{Contributions to $\kr$ at the test-point, $\log(T)=6.35$,
$\log(N_\rme)=23.0$. A04 abundances.}
\begin{tabular}{lc}
\hline
Skip & $\kr$\\%[2mm]
\hline
None & 15.29\\                  %15.75
H, He & 10.51\\                 %10.87
O     & 10.52\\                 %10.77
Fe group & 11.03\\                %11.16
\hline
\end{tabular}
\end{center}
\end{table}
\begin{table}
\begin{center}
\caption{Occupation probabilities and level populations for
hydrogenic oxygen at the test-point.}
%\vspace{3mm}
\begin{tabular}{ccccccc}
\hline
\multicolumn{1}{c}{}&
\multicolumn{2}{c}{OP}&
\multicolumn{2}{c}{OPAL}\\%[2mm]
%\hline
$n$ & $W(n)$ & {\tt POP}$(n)$ & $W(n)$ & {\tt POP}$(n)$\\%[2mm]
\hline
1 & 1.000 & 0.361 & 1.000 & 0.325\\
2 & 0.987 & 0.048 & 0.993 & 0.044\\
3 & 0.813 & 0.048 & 0.954 & 0.050\\
4 & 0.151 & 0.013 & 0.830 & 0.063\\
5 & 0.009 & 0.001 & 0.289 & 0.031\\
6 & 0.000 & 0.000 & 0.062 & 0.009\\
7 &       &       & 0.009 & 0.002\\
8 &       &       & 0.000 & 0.000\\
\hline
\end{tabular}
\end{center}
\end{table}
\begin{table}
\begin{center}
\caption{Ionization fractions for oxygen at the test-point ({\tt NE}$=-1$
for fully-ionized, {\tt NE}$=0$ for hydrogenic).}
\begin{tabular}{rcc}
\hline
{\tt NE} & OP & OPAL\\
\hline
$-$1 & 0.415 & 0.374\\
0  & 0.471 & 0.524\\
1  & 0.109 & 0.098\\
2  & 0.005 & 0.005\\
3  & 0.001 & 0.000\\
4  & 0.000\\
\hline
\end{tabular}
\end{center}
\end{table}
\begin{table}
\begin{center}
\caption{Rosseland-means at the test point.}
\begin{tabular}{lccl}
\hline
Mixture & OP & OPAL & \% difference\\
\hline
A04    & 15.29 & 14.99 & $+2.0$\\                   % 15.75 14.99 5.1
A04 less iron-group & 11.03 & 10.63 & $+3.7$\\[2mm] % 11.17 10.63 5.1
Difference & 4.26 & 4.38 & $-2.8$\\                 %  4.58  4.38 4.6
\hline
\end{tabular}
\end{center}
\end{table}
\clearpage
\subsection{Examination of the  Rosseland mean near the base of the convection zone}
In the region at and below the base of the convection zone the mean
opacities from OP and OPAL do not differ by more than 2.5\% (see Figure 7).
In view of the current interest in the opacity in that region, we
consider the results in some greater detail. 

In the BP04 model the base of the convection zone is at $\log(T)=6.338$,
$\log(\rho)=-0.735$. We consider the Rosseland-mean opacity at a
nearby test-point on the OP $(T,N_\rme)$ mesh,
$\log(T)=6.35$ and $\log(N_\rme)=23.0$ giving 
$\log(\rho)=-0.715$ for both BP04 and A04. Figure 8 shows $\log(\sigma(u))$
against $u=h\nu/(kT)$ calculated using BP04 abundances. The Rosseland
mean at the test-point is $\kr=19.35$  % 19.92
\mbox{cm$^2$ g$^{-1}$} with BP04 and 
15.29  % 15.75 
with A04.
\subsubsection{Contributing elements}
The code {\sc mixv.f} used to calculate Rosseland means has a ``skip''
facility: an element marked as ``skip'' is included in calculating
the mass density for a given electron density but its contribution
to $\sigma(u)$ is omitted. Table 2 shows the result of skipping: 
(a) hydrogen and helium; (b) oxygen; and (c) the iron-group elements.
It is seen that, for all three cases, substantial contributions
are skipped. We shall consider those three cases in greater detail.
\subsubsection{The contribution from hydrogen and helium}
OPAL and OP results for H and He are discussed in section 7.1 of
Paper III. The agreement is generally close but there is a region
in the vicinity of $\log(T)\sim 6$ where OPAL has larger populations
in the ground-states of H and of He$^+$ leading to larger OPAL opacities.
At our test-point, $\kr($OPAL$)$ for an H/He mixture ($X\simeq 0.7$,
$Y\simeq 0.3$) is about 5\% larger than $\kr($OP$)$, corresponding
to a difference of 2 or 3\% for the A04 mix.
\subsubsection{Oxygen}
Oxygen makes an important contribution to $\kr$ in the vicinity of
$\rcz$ because it has a high abundance and because about one half of the oxygen is in a hydrogenic
state (see Table 4) and the oxygen  Lyman lines and continuum come in a region
where the weighting function $F(u)$ is large (see Figure 8). The atomic
data for the hydrogenic stage should be quite accurate and those for
other high ionization stages should also be of good accuracy since,
for a highly ionized system, the electron--nuclear potentials will be large 
compared with electron--electron
potentials. The main uncertainties are likely to be in the level
populations, as discussed in Paper III.

We do not have OPAL occupation probabilities $W(n)$ for oxygen at the 
test-point, but we do 
have them for the rather similar case of carbon at $\log(T)=6.0$,
$\log(\rho)=-2.0$. From Table~1 of paper II we see that, compared with
OP, OPAL has larger occupation probabilities for more highly excited
states. From that table we obtain the approximate relation
\be W_{\rm OPAL}(n)\simeq W_{\rm OP}(0.716\times n)\ee
\begin{displaymath} (\mbox{carbon at}
\log(T)=6.0,\,\,\, \log(\rho)=-2)
\end{displaymath}
where we interpolate in $n$ as a continuous variable. We use (12) to
obtain estimates of $W_{\rm OPAL}(n)$ for hydrogenic oxygen. Table 3
gives populations $W(n)$ and populations {\tt POP}$(n)$ for hydrogenic
oxygen at the test-point, from
both OP and OPAL, and Table 4 gives the corresponding ionization
fractions, {\tt fion(NE)}. 

The Rosseland-mean cross-sections for oxygen are 
$\sigma_{\rm R}=4.011\times 10^{-4}$ % 4.086
atomic units ($a_0^2$) from OP and 
$3.721\times 10^{-4}$ with our estimates of OPAL occupation
probabilities. The corresponding opacities for the A04 mix
are 15.75 and 15.16 cm$^2$ g$^{-1}$.
The larger excited-state $W(n)$ from OPAL affect the
Rosseland means in two ways. Firstly, as noted in Paper II, they 
give a reduction the ground-state population and hence a reduction
in the strengths of the Lyman
absorption features. Secondly, they give increases in profiles
of lines having more highly excited states and reductions in
continuum absorption for dissolved lines (see Appendix A of Paper I).
Both effects lead to a reduction in Rosseland means. 

Our value of $\kr$ from OP is larger than that obtained using
our estimates of $W_{\rm OPAL}$ for oxygen by 4\%.  
\subsubsection{Other light elements}
It is shown in Paper III that, in the vicinity of $\rcz$, carbon
and sulphur have behaviours similar to that for oxygen, with
$\sigma($OP$)$ being a few per cent larger than $\sigma($OPAL$)$. 
It can be expected that the other light elements will have a similar
behaviour.
\subsubsection{The iron group}
It is seen from Table 2 that elements of the iron-group make
significant contributions to $\kr$ at our test-point. In the
vicinity of that point the dominant ionisation stages for
iron are oxygen-like and fluorine-like. The R-matrix atomic
data for the iron-group used by OP should be of rather good
accuracy. Table 5 gives values of $\kr$ at the test-point from
OP and OPAL. The OPAL values are obtained using tables from the
OPAL web site and the OPAL interpolation codes {\sc xztrin21.f}.
Results are given for:
\begin{enumerate}
 \item A04 abundances;
 \item A04 abundances less contributions from the iron group (number
fractions are set to zero for iron group elements, other number fractions
are unchanged except for H, for which the number fraction is increased
to ensure particle conservation). 
\end{enumerate}
It is seen that the OP contribution from the iron group is 2.8 \% 
smaller % was 4.6% larger !!
than the corresponding OPAL contribution.
\subsubsection{Conclusion}
We recall that Bahcall \etal (2004b) require $\kr$ to be 10\% larger than
$\kr($OPAL$)$ for \mbox{$ 2\times 10^{6} \lesssim T \lesssim 5\times 10^{6}$ K}\
($0.7$ to $0.4 R_{\sun}$). 
In that range we find that $\kr($OP$)$ never differs from $\kr($OPAL$)$ by more 
than 2.5\%.
The closeness of the agreement is partly fortuitous: compared with OP,
OPAL has larger contributions from H and He and from the iron-group, but
smaller contributions from oxygen. However, an increase to 10\% larger
than OPAL in the region \mbox{$2\times 10^6\lesssim T\lesssim 5\times 10^6$}
does not appear to be very plausible.

\section{Checks on accuracies of interpolations and of frequency mesh}
\label{chek}
Temperature and density indices $I$ and $J$ are defined by
\be \log(T)=0.025\times I,\,\,\,\log(\Ne)=0.25\times J.\ee
Four meshes are used:
\begin{enumerate}
\item fine mesh, `f', $\Delta I=\Delta J=1$;
\item medium mesh, `m', $\Delta I=\Delta J=2$;
\item coarse mesh, `c', $\Delta I=\Delta J=4$;
\item very coarse mesh, `C', $\Delta I=\Delta J=8$.
\end{enumerate}
The number of frequency points, equally spaced in the variable
$v$ (see Section \ref{mesh}), is taken to be equal to {\tt NTOT}. Normal OP production
work is done using the `m' mesh with even values of $I$ and $J$,
and {\tt NTOT}$=10000$. OPAL production work is done with 10000
equally spaced points in the variable $u=h\nu/(kT)$, $0.002 \le u \le 20$.
Checks on the accuracies of our integrations over frequency and
interpolations in $T$ and $\rho$ were made for a solar mix.

Firstly, check calculations were made using the `C' mesh with {\tt NTOT}$=10000$
and $20000$. It was found that use of the finer frequency mesh
({\tt NTOT}$=20000$) never led to a change in the Rosseland-mean
of more than 0.1\%.

Then, `C' calculations were made for odd values of $I$ and $J$, so that
none of the `C' mesh-points coincided with `m' mesh-points. Using 
values of Rosseland-means for the `m' points,
interpolations to the
`C' points were made using the code {\sc opfit.f} (Seaton, 1993). 
For $\log(T)\ge 3.73$ and
$\log(R)\le -1.0$ all differences are less than 0.6\%.
Larger differences occur only in two regions: small $\log(T)$,
where $\kr$ is varying rapidly as a function of $T$ (see
Fig. 1) --- the worst case is 1.6\% at ($\log(T)=3.525$,
$\log(R)=-7.2$); and large $\log(R)$, close to the upper
boundary for {\sc opfit} interpolations (worst cases are
3.1\% at ($\log(T)=6.525$, $\log(R)=-0.530$) and 5.8\% at 
($\log(T)=7.25$, $\log(R)=-0.531$)). 

Bahcall \etal (2004) report that the use of OPAL
opacities and different interpolation schemes yield opacity
values in the solar-centre region differing by as much as 4\%.
In view of the importance of obtaining accurate opacities for
that region, we have made further checks.  For $6.0 \le \log(T) \le 8.0$
we have made  `m' mesh calculations with  {\tt NTOT} equal both 10000 and
30000. For all densities considered in the OP work ($\log(\Ne)=15.5$
to 23.5 for $\log(T)=6.0$, $\log(\Ne)=22.0$ to 29.5 for $\log(T)=8.0$)
we find the largest differences in single-element Rosseland-means,
for {\tt NTOT}$=10000$ and 30000, to be about 4\%. Such
comparatively large differences occur only for cases with deep
minima in the monochromatic opacities. For typical mixtures
such minima get filled-in by contributions from other
elements and the sensitivities to {\tt NTOT} become much smaller. For a solar-type
mix, $6.0 \le \log(T) \le 8.0$ and all densities, the differences
between $\kr$ for {\tt NTOT}$=10000$ and 30000 are never larger than
0.1\%.
 
As a further check on interpolations we have made calculations
using both the `m' and `f' meshes with {\tt NTOT}$=10000$. We used
{\sc opfit.f} to make calculations for a model with a fine mesh in
$\log(T)$ and densities such as to give $\log(R)=-1.75$,
roughly corresponding to a solar-centre model. For 
$6.2 \le \log(T) \le 7.5$, values of $\kr$ from the `m' and
`f' calculation never differ by more than 0.4\%: for the 
vicinity of the base of the convection zone they never differ
by more than 0.1\%.
\section{Data archives}
\subsection{Outer-shell atomic data}
The OP outer-shell atomic data are archived in TOPbase, which is 
part of the OP  archive at the CDS (Centre
de Donn\'{e}es de Strasbourg)\footnote{
\tt http://cdsweb.u-strasbg.fr/topbase}. The original version of
TOPbase is described by Cunto \etal (1993).
\subsection{Atomic data from {\sc autostructure}}
Both inner- and outer-shell data from {\sc autostructure} have been
archived according to the Atomic Data and Analysis Structure (ADAS)
{\it adf38} and {\it adf39} data formats (Summers 2003).
We have included  LS and IC final-state resolved 
photoexcitation--autoionization and
photoionization data, summed over final channel angular momenta
for both inner- and outer-shells. 
This will enable the data to be used in the future for other
purposes such as the collisional--radiative modelling of photoionized plasmas.
Further details of the data archival may be found in Paper II.
\subsection{Monochromatic opacities and computer codes}
Codes for the calculation of Rosseland-mean opacities  and 
radiative accelerations for any chemical mixtures, 
temperatures and densities are described  by Seaton (2005).
A {\tt tar} file containing the codes and all required monochomatic
opacities is available for download from the TOPbase web site at CDS. Since
the file size is a little under 700Mb, it is also available on 
a CD\footnote{Requests to: {\tt Claude.Zeippen@obspm.fr}}.
\subsection{OPserver}
Rosseland mean opacities and their derivatives can also be computed online 
for any required chemical mixture using the OPserver at the Ohio Supercomputer
Center, Columbus, Ohio, USA (Mendoza \etal, 2005). It can be accessed 
{\em via} the TOPbase web site at CDS.  
\section{Summary}
OP opacities have been up-dated by inclusion of all contributing
inner-shell processes and some further refinements have been made.

Comparisons with OPAL show agreement generally to within 5 to 10\%.

Using earlier estimates of metal abundances, good agreement was
obtained between computed solar models and results obtained
from helioseismology for $\rcz$ and other data. With recent downward revisions in metal
abundances that agreement is destroyed. It is shown by Bahcall \etal
(2004b) that agreement would be restored if opacities in the region
just below the base of the convection zone,
\mbox{$2\times 10^6 \lesssim T \lesssim 5\times 10^6$ K}
($0.7$ to $0.4 R_{\sun}$), were 10\% larger than those from OPAL . In that 
region we find $\kr($OP$)$ never to be larger than $\kr($OPAL$)$ by
more than 2.5\%. Results for opacities in the region near the base of 
the convection zone have been discussed in some detail.

Studies have been made of the accuracies of the OP integrations over
frequency, used to obtain Rosseland means, and of the OP procedures
used for interpolations in temperature and density. It is concluded
that the OP Rosseland-means should be numerically correct to
within 1\% or better.

\section*{Acknowledgements}
We thank Dr N. Grevesse for providing the table of abundances
which we refer to as A04, and Dr A. M. Serenelli for providing
the table of abundances for BP04. 
We thank Dr J. N. Bahcall for advice and comments.
FD is grateful to the INSU/CNRS for support during visits to Meudon.
\bsp

\clearpage
\begin{center}
\epsfig{file=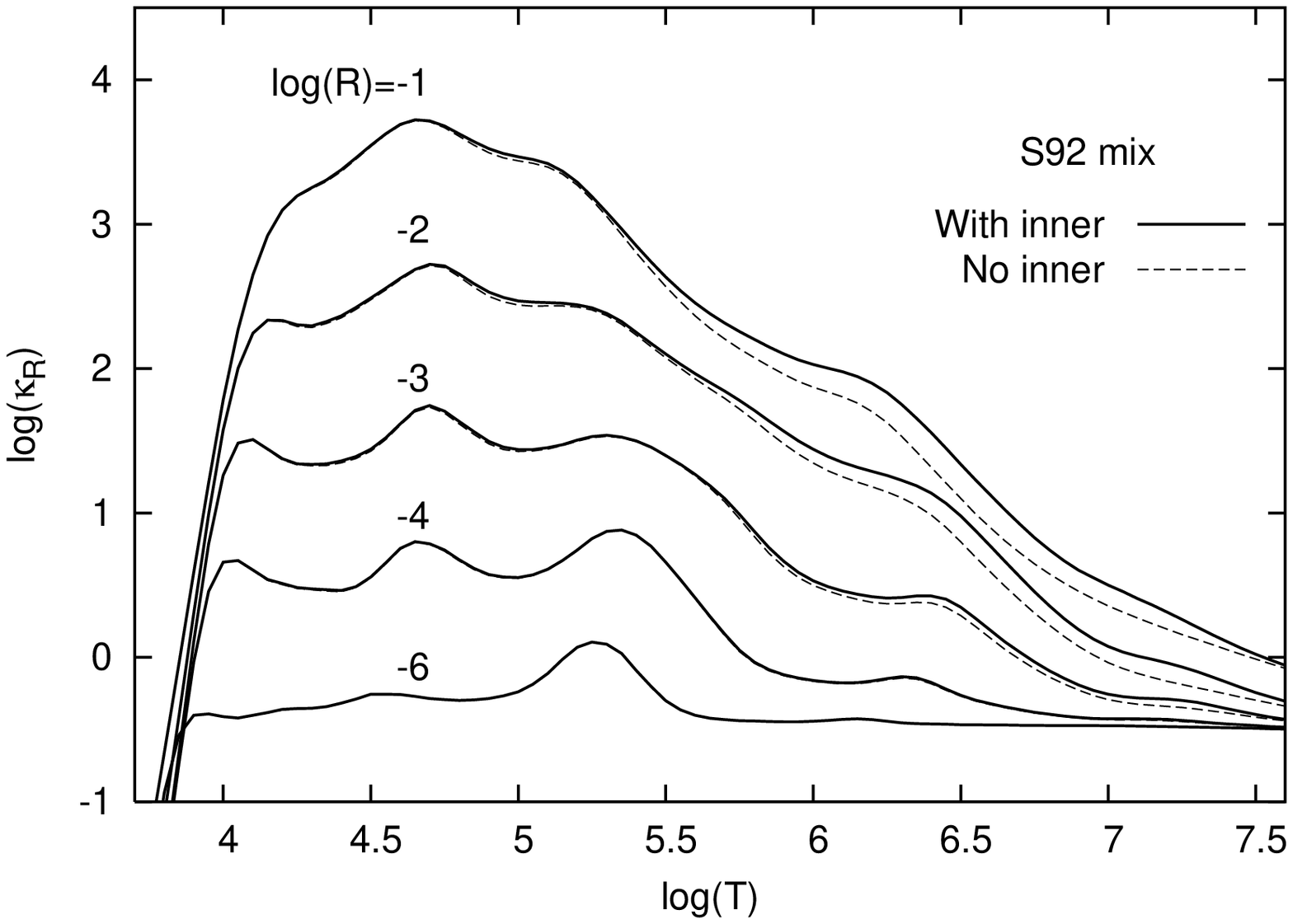,angle=270,width=10cm}
\end{center}
\vspace{2mm}
{\bf Figure 1.} Rosseland-mean opacities from OP for S92 mix, with and
without inner-shell contributions.
\vspace{10mm}
\begin{center}
\epsfig{file=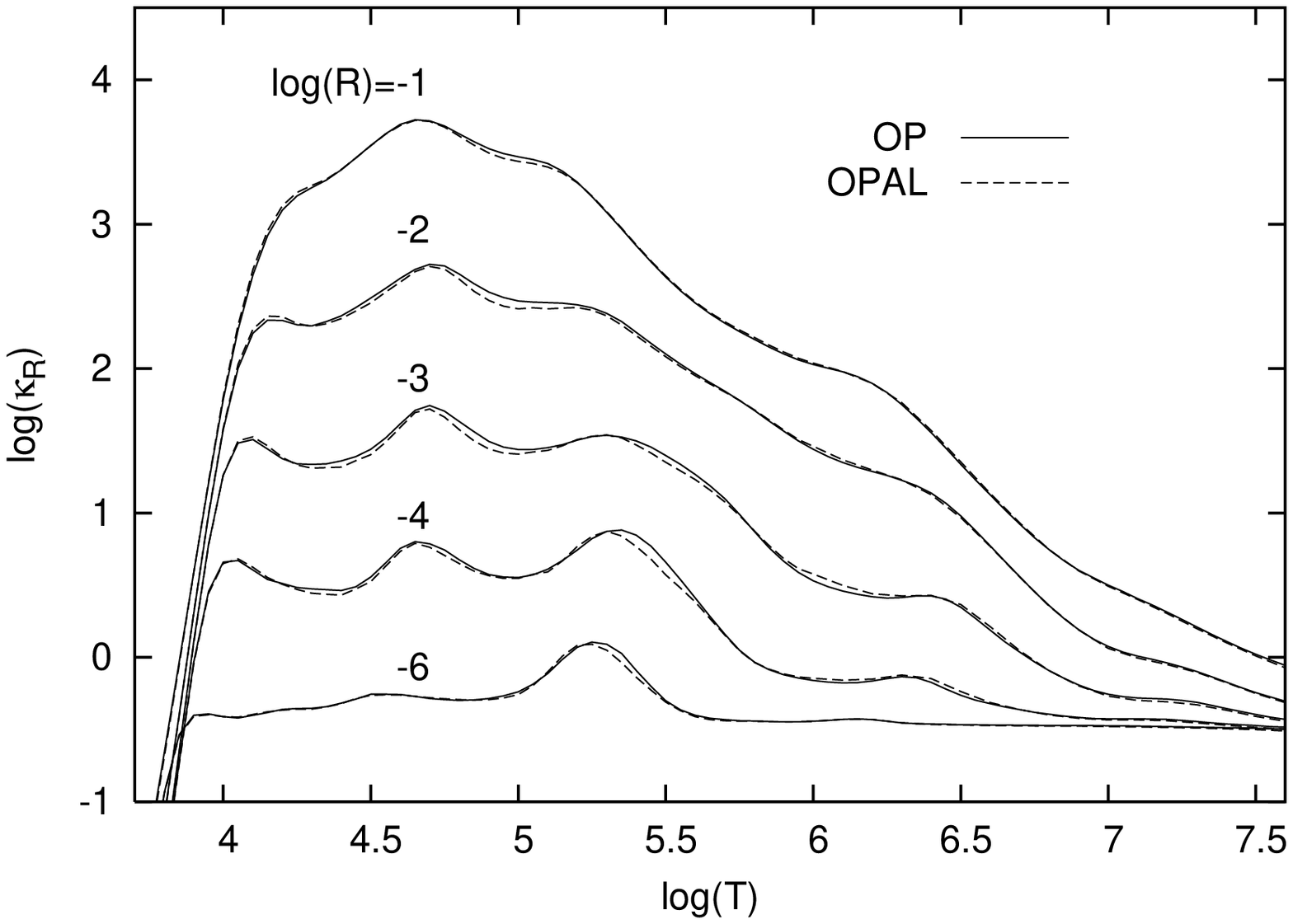,angle=270,width=10cm}
\end{center}
\vspace{2mm}
{\bf Figure 2.} Rosseland-mean opacities from OP and OPAL for the S92 mix.
\clearpage
.
\vspace{30mm}
\begin{center}
\epsfig{file=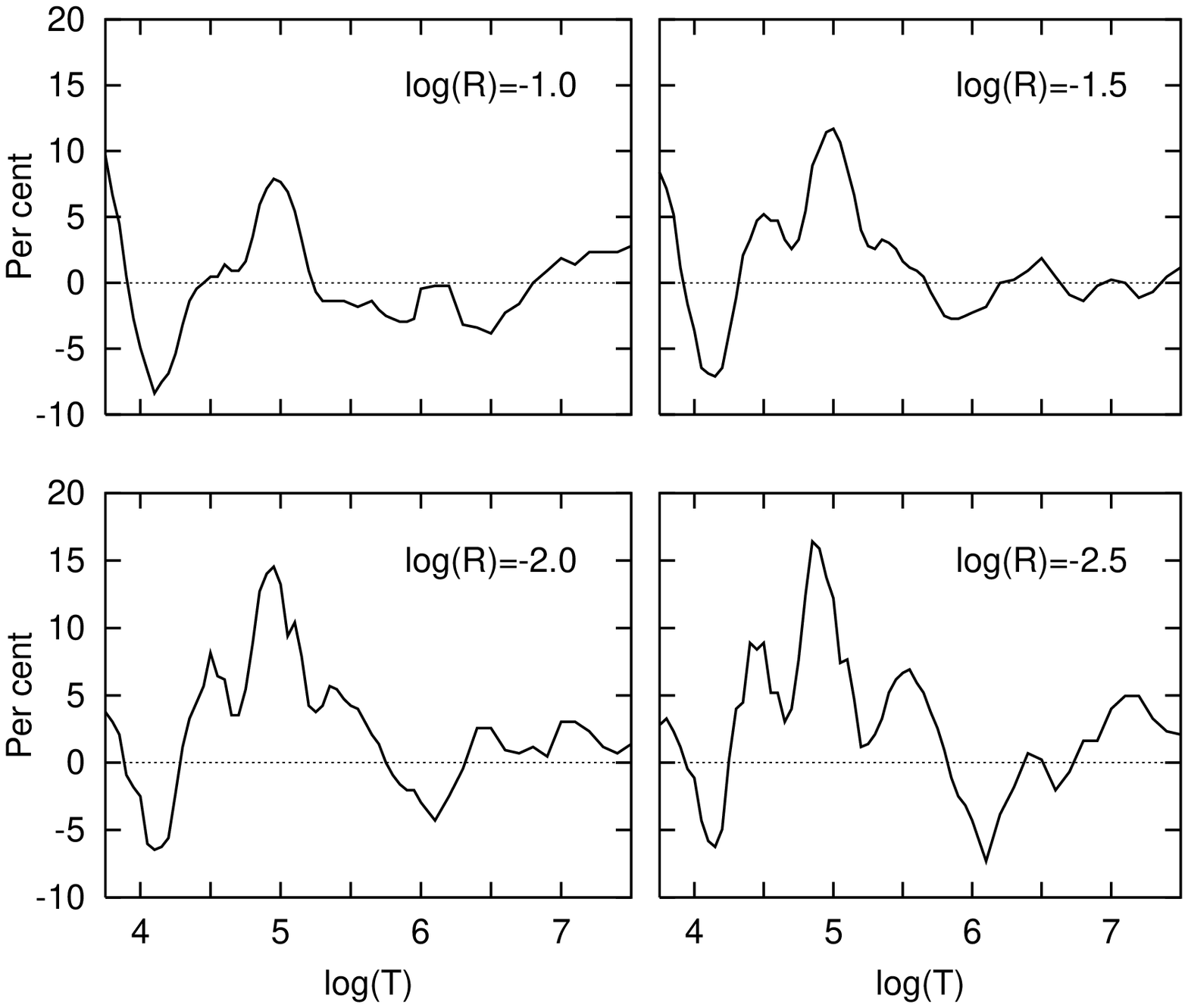,angle=270,width=8cm}
\end{center}
\vspace{2mm}
{\bf Figure 3.} Percentage differences, (OP--OPAL), for the S92 mix: $\log(R)=-1$
to $-2.5$.
\vspace{35mm}
\begin{center}
\epsfig{file=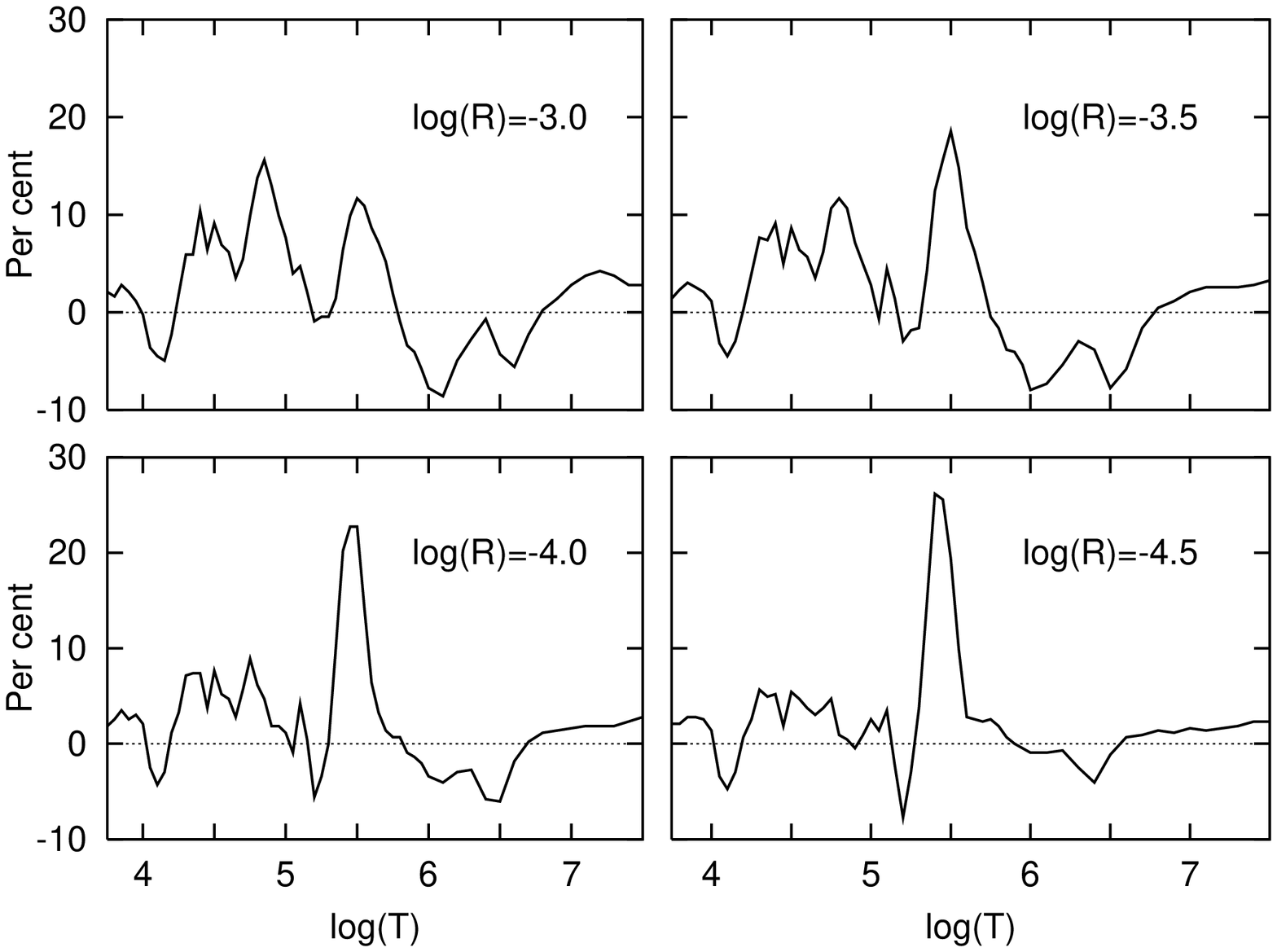,angle=270,width=7.46cm}
\end{center}
\vspace{2mm}
{\bf Figure 4.} As Figure 3, for $\log(R)=-3$ to $-4.5$.
\clearpage
\begin{center}
\epsfig{file=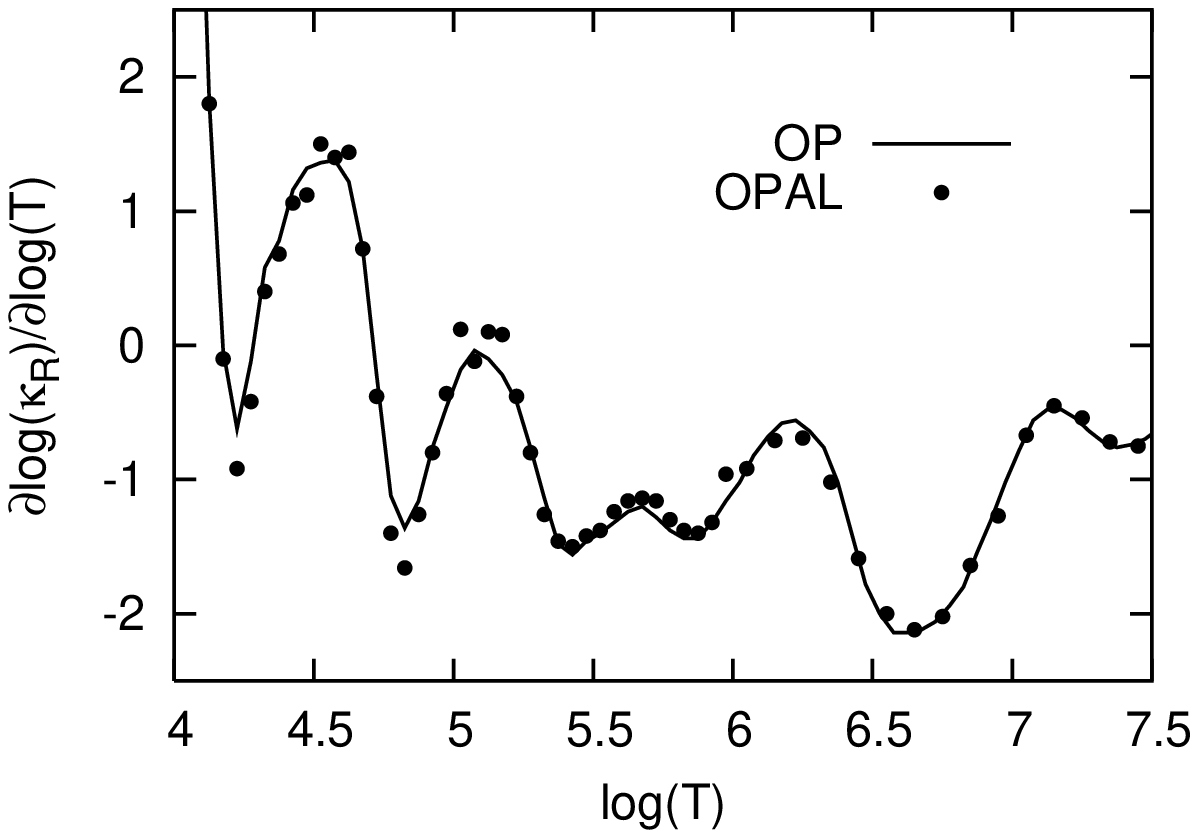,angle=270,width=10cm}
\end{center}
\vspace{2mm}
{\bf Figure 5.} Values of $\partial\log(\kr)/\partial\log(T)$ (at
constant $R$) approximated using first derivatives. From OP
and OPAL for $\log(R)=-2$.
\clearpage
\begin{center}
\epsfig{file=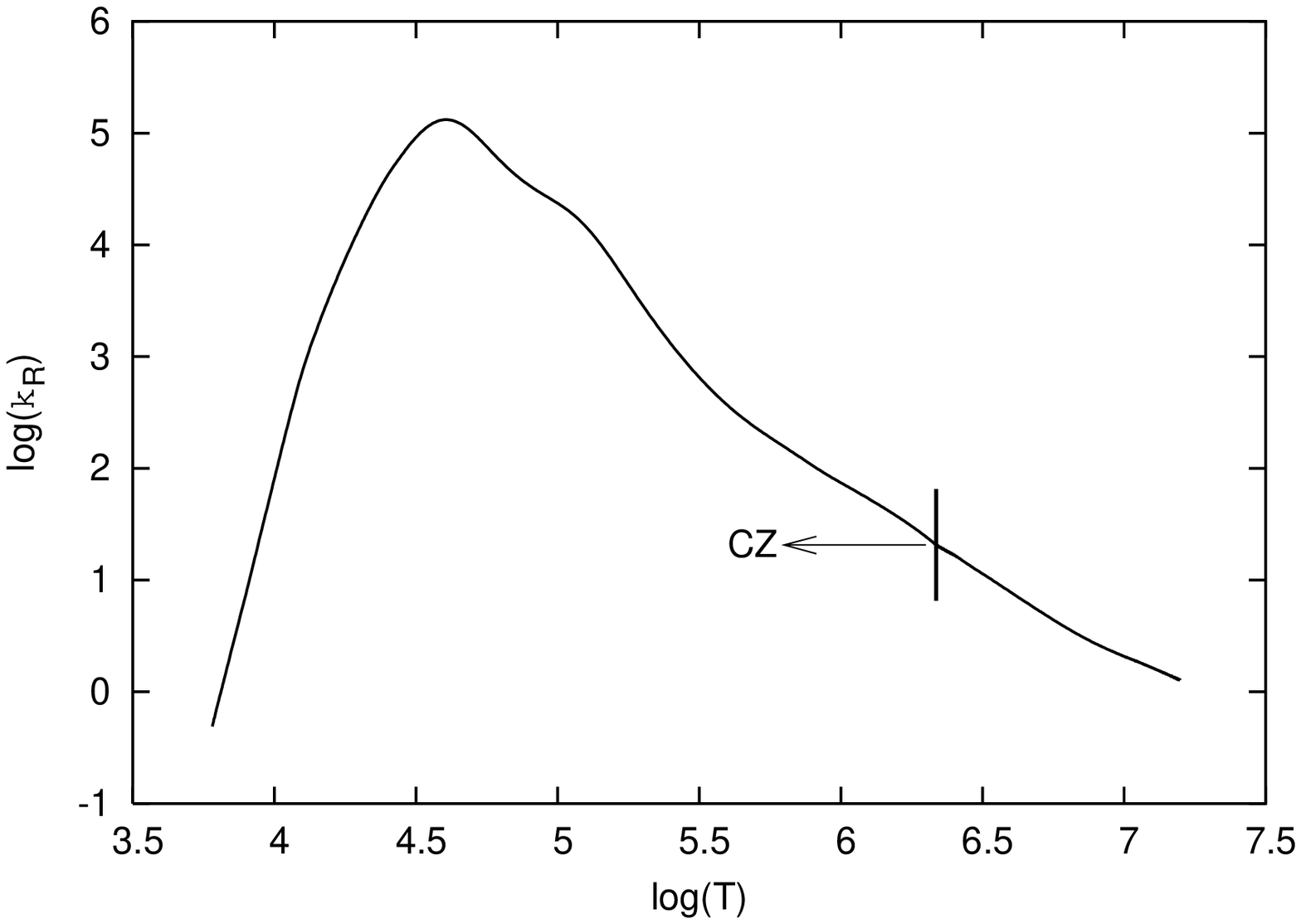,angle=270,width=10cm}
\end{center}
\vspace{2mm}
{\bf Figure 6.} Rosseland-mean opacity from OPAL as used in the
solar model BP04 of Bahcall and Pinsonneault (2004). The convection
zone is the region indicated with $\log(T) \lesssim 6.34$.
\begin{center}
\epsfig{file=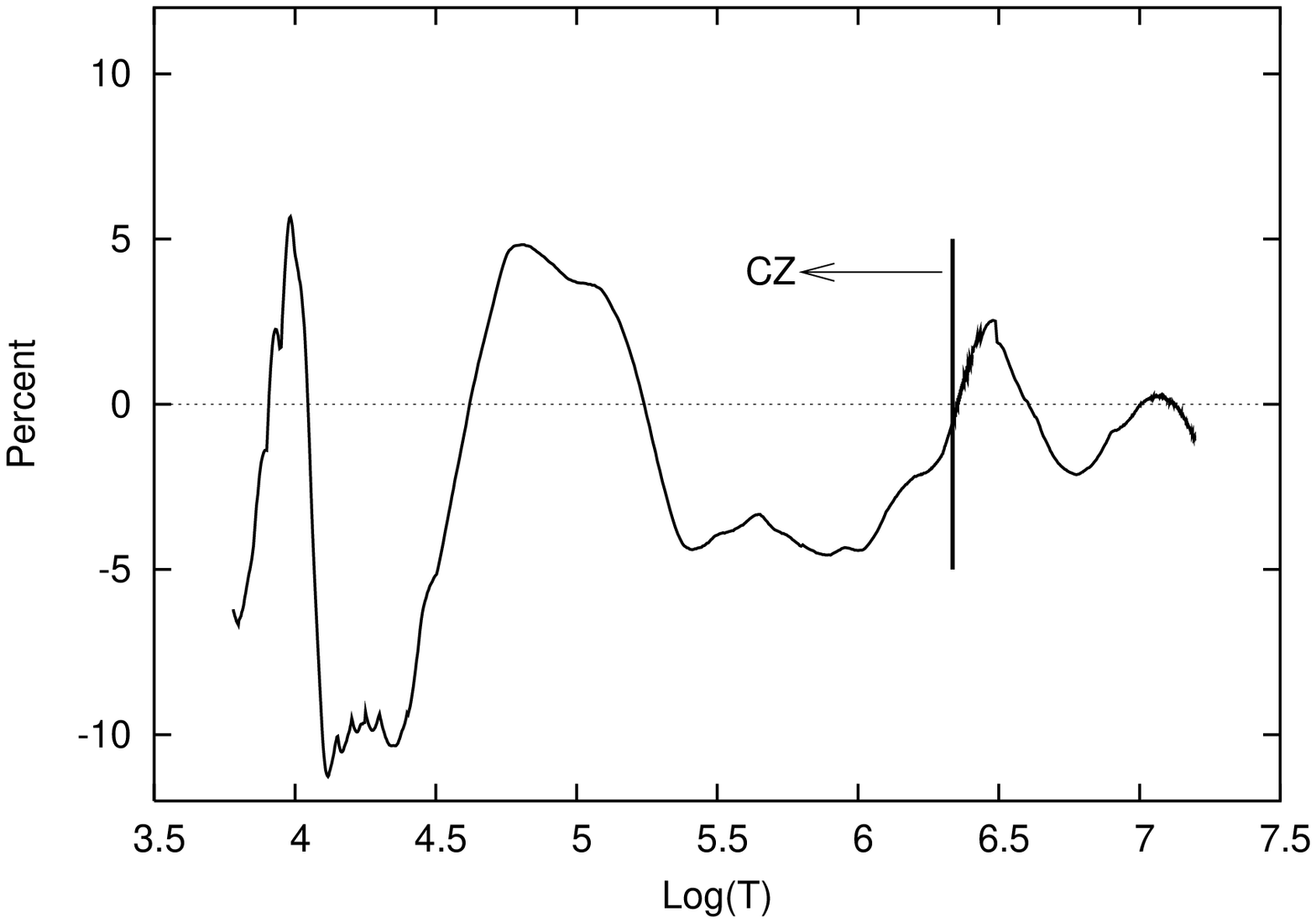,angle=270,width=10cm}
\end{center}
\vspace{2mm}
{\bf Figure 7.} Percentage differences, (OP--OPAL), between opacities
for a solar model from Bahcall and Pinsonneault (2004).
The convection zone is the region with $\log(T) \lesssim 6.34$.
\clearpage
\begin{center}
\epsfig{file=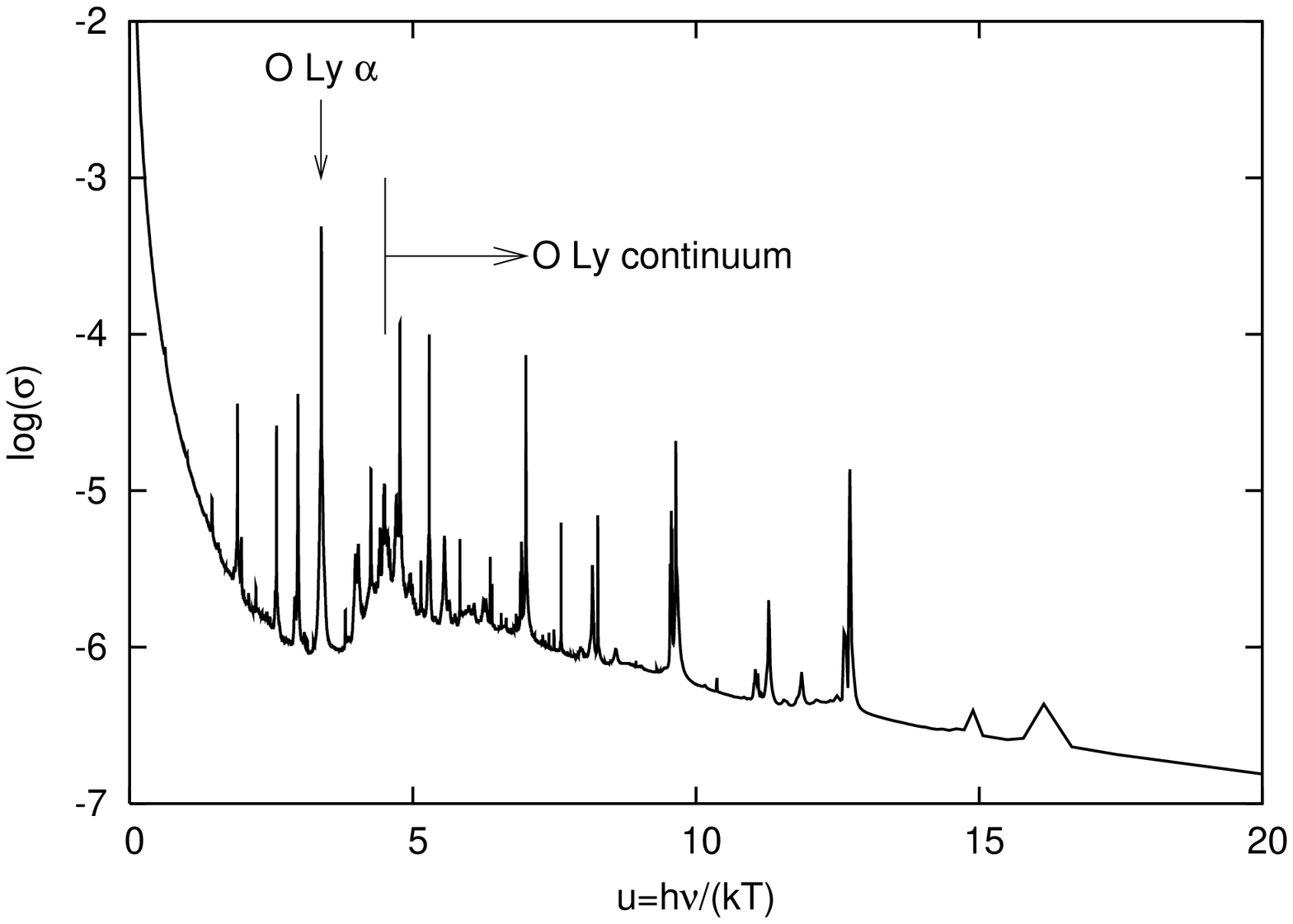,angle=270,width=10cm}
\end{center}
\vspace{2mm}
{\bf Figure 8.} The monochromatic opacity for the BP04 mix at
$\log(T)=6.35$, $\log(\Ne)=23.0$. Cross-section, $\sigma$, in
atomic units, $\pi a_0^2$.
\label{lastpage}
\end{document}